\documentclass[sigconf,nonacm]{acmart}

\AtBeginDocument{%
  \providecommand\BibTeX{{%
    \normalfont B\kern-0.5em{\scshape i\kern-0.25em b}\kern-0.8em\TeX}}}

\setcopyright{acmcopyright}
\copyrightyear{2023}
\acmYear{2023}
\acmDOI{XXXXXXX.XXXXXXX}

\acmConference[ICFNDS23]{the 7th International Conference on Future Networks and Distributed Systems}{December 21--22,
  2023}{Dubai, UAE}
\acmPrice{15.00}
\acmISBN{978-1-4503-XXXX-X/18/06}

\usepackage[most]{tcolorbox}

\newtcolorbox{myquote}[1][]{%
    colback=black!5,
    colframe=black!5,
    notitle,
    sharp corners,
    enhanced,
    breakable,
    }
    
\begin{document}

\title{Data Trading and Monetization: Challenges and\\ Open Research Directions}


\author{Qusai Ramadan}
\affiliation{%
  \institution{Institute of Software Engineering}
  \institution{University of Koblenz}
  \country{Germany}}
\email{qramadan@uni-koblenz.de}

\author{Zeyd Boukhers}
\affiliation{%
  \institution{Fraunhofer Institute for Applied \\ Information Technology FIT, and }
  \institution{Uniklinik K{\"o}ln}
  \country{Germany}
}
\email{zeyd.boukhers@fit.fraunhofer.de}

\author{Muath AlShaikh}
\affiliation{%
  \institution{College of Computing and Informatics}
   \institution {Saudi Electronic University}
  \country{Kingdom of Saudi Arabia}}
\email{m.alshaikh@seu.edu.sa}

\author{Christoph Lange}
\affiliation{%
  \institution{Fraunhofer Institute for Applied \\ Information Technology FIT, and }
  \institution{RWTH Aachen University}
  \country{Germany}
}
\email{christoph.lange-bever@fit.fraunhofer.de}

\author{Jan J{\"u}rjens}
\affiliation{%
  \institution{University of Koblenz, and} 
  \institution{Fraunhofer-Institute for Software and Systems Engineering ISST}
  \country{Germany}}
\email{http://jan.jurjens.de}


\begin{abstract}
  Traditional data monetization approaches face challenges related to data protection and logistics. In response, digital data marketplaces have emerged as intermediaries simplifying data transactions. Despite the growing establishment and acceptance of digital data marketplaces, significant challenges hinder efficient data trading. As a result, few companies can derive tangible value from their data, leading to missed opportunities in understanding customers, pricing decisions, and fraud prevention. In this paper, we explore both technical and organizational challenges affecting data monetization. Moreover, we identify areas in need of further research, aiming to expand the boundaries of current knowledge by emphasizing where research is currently limited or lacking. 
\end{abstract}




\maketitle

\section{Introduction}\label{sec:intro}
The amounts of data being created, captured, and replicated are continuously increasing at a fast pace. According to a recent study, the projected growth of annual data on a global scale indicates a substantial rise from 33 zettabytes in 2018 to an estimated 175 zettabytes by the year 2025~\cite{rydning2018digitization}. The availability of data and advanced technologies offer substantial potential for creating value. Therefore, within the business landscape, data has transitioned from being merely an enabler of products to becoming valuable products in and of themselves, serving as strategic resources for companies~\cite{spiekermann2019data}. For example, it is predicted that the European Union's data economy will attain a value of €829 billion, accompanied by a notable expansion in the workforce of data professionals, expected to reach 10.9 million by 2025~\cite{rydning2018digitization}. 

In situations where organizations lack efficient amounts and high-quality data to effectively execute or enhance their processes and services (which is the case in many SMEs) they may seek to acquire data from other organizations through purchasing arrangements. Traditionally, data monetization involves selling or licensing data directly to third parties. However, this approach often raised privacy concerns and required transferring large volumes of data, which could be time-consuming and costly~\cite{koutroumpis2020markets}. As a result, in recent times, a multitude of digital data marketplaces and platforms have emerged, serving as intermediaries between data providers and consumers. Data marketplaces represent a recent development in the market, with recent instances emerging as a means to facilitate industrial data trading and monetization.

Despite the growing establishment and acceptance of data marketplaces in industry (with more than 70\% of enterprises in the EU having already acquired external data), substantial obstacles persist that hinder data trading~\cite{koutroumpis2020markets}. In this paper, we shed light on the current technical and organizational challenges that affect data monetization. We also identify open research directions to push the boundaries of existing knowledge by highlighting areas where research is limited or lacking.  

This paper is organized as follows: in Section~\ref{sec:background} we provide the necessary background. Section~\ref{sec:lit} provides the state of the art. In Section~\ref{sec:challenges} we explore the challenges that affect data monetization, and we propose future research directions toward addressing the identified challenges. In Section~\ref{sec:discussion} we discuss the limitations of our research findings. Finally, Section~\ref{sec:conclusion} concludes this paper.   

\section{Background}\label{sec:background}
Online data marketplaces have evolved through different phases and in different categories. Initially, data catalogues focused on open data, providing search and browse mechanisms for metadata and links to data sets. Then, government and industrial marketplaces facilitated secure data exchange and data sovereignty. In later phases, personal data markets and open science marketplaces emerged\footnote{\url{https://marketplace.eosc-portal.eu/}, Accessed: (14 September 2023)}, along with Sensor Data Marketplaces for IoT data\footnote{\url{https://rubygarage.org/blog/big-data-marketplaces}, Accessed: (14 September 2023)}. Three distinct directions have been deliberated and implemented concerning the architecture of data marketplaces:

\smallskip
\noindent\textbf{Centralized data marketplaces.} The centralized marketplace design acts as a standard multi-sided platform, attracting various participants from the data ecosystem and providing facilitation services through a technological platform, enabling the platform operator to benefit from positive participation~\cite{thomas2016big}. 

A centralized design prioritizes larger data suppliers as costs decrease with scale. However, creating and maintaining diverse raw data records is challenging. Ideally, the marketplace should possess metadata for all data records to enable future trades~\cite{ramachandran2018towards}. In addition to the expensive maintenance of this type of platform, the distribution of exchange volumes and profits will heavily favour the larger entities that provide high volumes of data at a fixed cost of metadata~\cite{spiekermann2019data}.  A centralized platform can enforce specific entrance policies and fees, which creates a clear boundary. A centralized design without boundaries and enforcement is more suitable for non-private data. However, for private data, there is a risk of misuse after the exchange. 

\smallskip
\noindent\textbf{Decentralized data marketplaces.} Distributed Ledger Technologies (DLTs) and blockchains have sparked interest in decentralized marketplace designs. Initially applied in virtual currency markets~\cite{evans2014economic}. In the decentralized design, market clearing, and transparency are improved, resulting in a shift of taxonomic tagging responsibilities from the centralized marketplace to individual data suppliers~\cite{upadhyay2021blockchain,catalini2020some}. Each data provider must maintain a data taxonomy that annotates records with common standards for searching, aggregating, analyzing, and reselling data. 

A decentralized marketplace has similar features to a centralized marketplace but overcomes some of its drawbacks. The key advantage is that trades can be executed and verified directly by participants. Additionally, proof of provenance is decentralized and can be verified independently~\cite{koutroumpis2017unfulfilled,swan2015blockchain}.  In contrast to centralized approaches, a decentralized marketplace tackles three crucial challenges: Firstly, it addresses the issue of data privacy by recording trades in a public ledger and authenticating transactions, eliminating privacy obstacles in trading~\cite{zyskind2015decentralizing}. Secondly, it distinguishes between data asset holders and original creators, granting creators the right to control future access and usage if desired, thereby ensuring ownership and control over data assets~\cite{koutroumpis2017unfulfilled}. Lastly, it enables an open and transparent exchange environment, allowing for easy search based on unique characteristics~\cite{bhujel2022survey}.

\smallskip
\noindent\textbf{Federated data marketplaces.} Federated data marketplaces allow organizations to collaborate on data analysis without sharing the underlying raw data~\cite{xu2021fed}.  Federated data marketplaces differ from decentralized data marketplaces regarding data governance and control. In decentralized data marketplaces, data is typically stored and managed using distributed ledger technology (e.g., blockchain), where data owners have direct control over their data and can grant access to specific parties based on predefined rules and smart contracts. In contrast, federated data marketplaces rely on federated learning~\cite{li2020blockchain} or other privacy-preserving techniques~\cite{hynes2018demonstration} to enable collaborative data analysis. Data remains in the possession of the original owners, and analysis is performed on local data sets. Aggregated results or model updates are then shared among the participants while preserving data privacy. This way, organizations can unlock the value of their data assets while maintaining control over data privacy and compliance.  

\section{State of the Art}\label{sec:lit}
We conducted desk research to provide an overview of the current status of data marketplaces. In this section, we discuss the most related industrial-oriented research projects, initiatives, and standards. Our research deliberately focuses on recent developments in the field of data trading within the European Union (EU). We have limited the scope of our study to this specific region due to the EU's prominent role in shaping policies and frameworks for secure data exchange. For instance, following the introduction of the EU's 2030 vision, which seeks to establish a unified European Data Market facilitating seamless data exchange across sectors and member states, numerous research projects have been funded by the EU Commission towards realizing the EU vision. 

\subsection{Projects and Initiatives}
In the following, we list the most related industrial-oriented projects and initiatives that have been developed within the EU towards facilitating data trading and monetization. 

\smallskip
\noindent\textbf{AI4EU.} This initiative constructed an on-demand platform for AI within Europe, with the objective of reducing barriers to innovation, fostering technology transfer, and facilitating the growth of startups and SMEs across all sectors\footnote{\url{https://www.ai4europe.eu/}, accessed: 14/09/2023)}. Its goals included promoting shared research, ensuring long-term sustainability, expanding the availability of AI assets within its ecosystem, and ensuring interoperability with existing AI and data components. Additionally, AI4EU established the Ethical Observatory to guarantee adherence to human-centric AI values, European data privacy regulations, and AI systems' transparency.

\smallskip
\noindent\textbf{AEQUITAS.} This project, funded by the European Union in 2022, has set out to establish a controlled experimentation environment with the aim of assisting AI producers in enhancing their awareness of the discrimination that could result from AI systems\footnote{\url{https://www.aequitas-project.eu/consortium}, accessed: 14/09/2023)}. Specifically, AEQUITAS proposes the design of a controlled experimentation environment that caters to both developers and users, enabling them to conduct controlled experiments for various purposes. These include detecting and mitigating bias in AI systems. This project provides an experimentation environment to generate synthetic data sets with diverse characteristics that influence fairness, enabling testing in laboratory settings.

\smallskip
\noindent\textbf{EUH4D.} This project, initiated in 2020 by the European Union, aimed to establish a federation of Big Data Digital Innovation Hubs (DIHs) across Europe\footnote{\url{https://euhubs4data.eu/overview/}, accessed: 14/09/2023)}. The project's main objective is to develop a European catalog of data sources and federated data-driven services, make this offer accessible at the regional level for European SMEs and start-ups, and foster cross-border data-driven experimentation through data sharing and interoperability. To achieve this objective, the project established an initial ecosystem of 12 Big Data DIHs from the four EU poles. Furthermore, the project aimed to engage SMEs, start-ups, and web entrepreneurs, conducting cross-border experiments and involving a significant number of companies in the data-driven innovation program. Starting with 12 EU regions in nine countries, EUHubs4Data planned to expand its reach to over 20 regions in 14 countries, establishing an ecosystem for data-driven innovation in Europe.

\smallskip
\noindent\textbf{DATAPorts.} This project, initiated in 2020 and funded by the European Union, aimed to transition European seaports from connected and digital entities to smart and cognitive ones\footnote{\url{https://dataports-project.eu/}, accessed: 14/09/2023)}. It provided a secure environment for aggregating and integrating data from diverse sources within digital ports, owned by different stakeholders. This was achieved through the design, implementation, and operation of the Cognitive Ports Data Platform, which interconnected existing digital infrastructures within seaports, established trusted data sharing and trading policies, and offered advanced data analytic services. The project involved deploying the platform in two European seaports, addressing local constraints, and conducting a global use case to target inter-port objectives. The broader implications of the DataPorts project extend to seaports across Europe, enhancing the trustworthiness, reliability, and efficiency of business operations and reinforcing the European Single Market.

\smallskip
\noindent\textbf{MUSKETEER.} An EU-funded project initiated in 2018, aimed to create a validated, federated machine learning platform capable of addressing privacy preservation, scalability, and efficiency challenges for real-world applications in the field of smart manufacturing and healthcare \footnote{\url{https://musketeer.eu/}, accessed: 14/09/2023)}.  The key objectives of MUSKETEER include creating machine learning models suitable for privacy-preserving scenarios, establishing robust security measures against external and internal threats, developing a standardized and expandable architecture, and introducing a rewarding model to monetize data sets based on their real value.

\smallskip
\noindent\textbf{TRUSTS.} This EU-funded project, running from 2020 to 2022, aimed to establish trust and confidence in data markets by developing an independent and federated platform based on insights gained from previous national projects\footnote{\url{https://www.trusts-data.eu/}, accessed: 14/09/2023)}. A primary focus of the project was the thorough examination of legal and ethical aspects throughout the entire data valorization chain. With the objective of creating a GDPR-compliant data marketplace for both personal and non-personal data, TRUSTS enhanced existing marketplaces and showcased the potential of the TRUSTS Platform through various use cases in the corporate business data sector.  

\subsection{Standards}
In the following, we discuss the most related reference architectures and standards that have been developed within the EU to facilitate data trading.

\smallskip
\noindent\textbf{International Data Spaces Reference Architecture Model (IDS-RAM).} The International Data Spaces Association (IDSA) has defined the IDS Reference Architecture Model (IDS-RAM) and agreements to establish trusted virtual data spaces~\cite{otto2019international}. The central component for secure data exchange is the IDS connector, which acts as a trusted security gateway. It enables direct transmission of data from certified data spaces, allowing the original data provider to maintain control and set usage conditions~\cite{pettenpohl2022international}. The connector encapsulates data within a virtual container, ensuring compliance with mutually agreed-upon terms.

It is noteworthy that IDS operates as an architecture, not a platform, enabling seamless integration with other systems to universally implement the concept of data spaces across various domains\footnote{\url{https://www.fraunhofer.de/content/dam/zv/en/fields-of-research/industrial-data-space/whitepaper-industrial-data-space-eng.pdf}, accessed: 14/09/2023)}. To establish formal trust within this architecture, The IDSA has developed a certification concept for IDS components and participants' operational environments. Independent evaluation facilities ensure transparent and trust mechanisms. Optional components, such as the Clearing House, Vocabulary Provider, Broker, and App Store, prepare the system for utilization as a data marketplace\footnote{\url{https://docs.internationaldataspaces.org/ids-knowledgebase/v/ids-ram-4/}, accessed: 14/09/2023)}. 

\smallskip
\noindent\textbf{Gaia-X.} The Gaia-X initiative was initiated in Europe and is now supported and led by representatives from politics, business, and science in France, Germany, and other European partners\footnote{\url{https://www.gaia-x.eu/}, accessed: 19/09/2023)}. It aims to promote interoperability and establish data and infrastructure ecosystems aligned with European values and standards. Its primary goal is to enhance the development of a federated, trusted, and user-friendly digital ecosystem. The project addresses various challenges, including decentralized processing locations, diverse technology stacks, limited transparency and sovereignty over stored and processed data and infrastructure, ambiguity regarding applicable jurisdiction, sector-specific data spaces and ontology gaps, insufficient accessibility of widely accessible application programming interfaces (APIs), multiple stakeholders, and limited accessibility to existing data and infrastructure services~\cite{braud2021road,niebel2022gaia}.

Gaia-X does not aim to create a competing product but rather seeks to network different elements through open interfaces and standards, linking data and fostering innovation. By leveraging existing approaches, Gaia-X aims to offer an alternative data ecosystem approach to the dominant global platforms~\cite{tardieu2022role}. It establishes its own data sharing and storage services, ensuring digital sovereignty for data owners while providing a foundation for smart services and innovative business processes.


\section{Challenges and Open Research Directions}\label{sec:challenges}
\begin{figure*}[t]
	\centering
	\includegraphics[width=\textwidth]{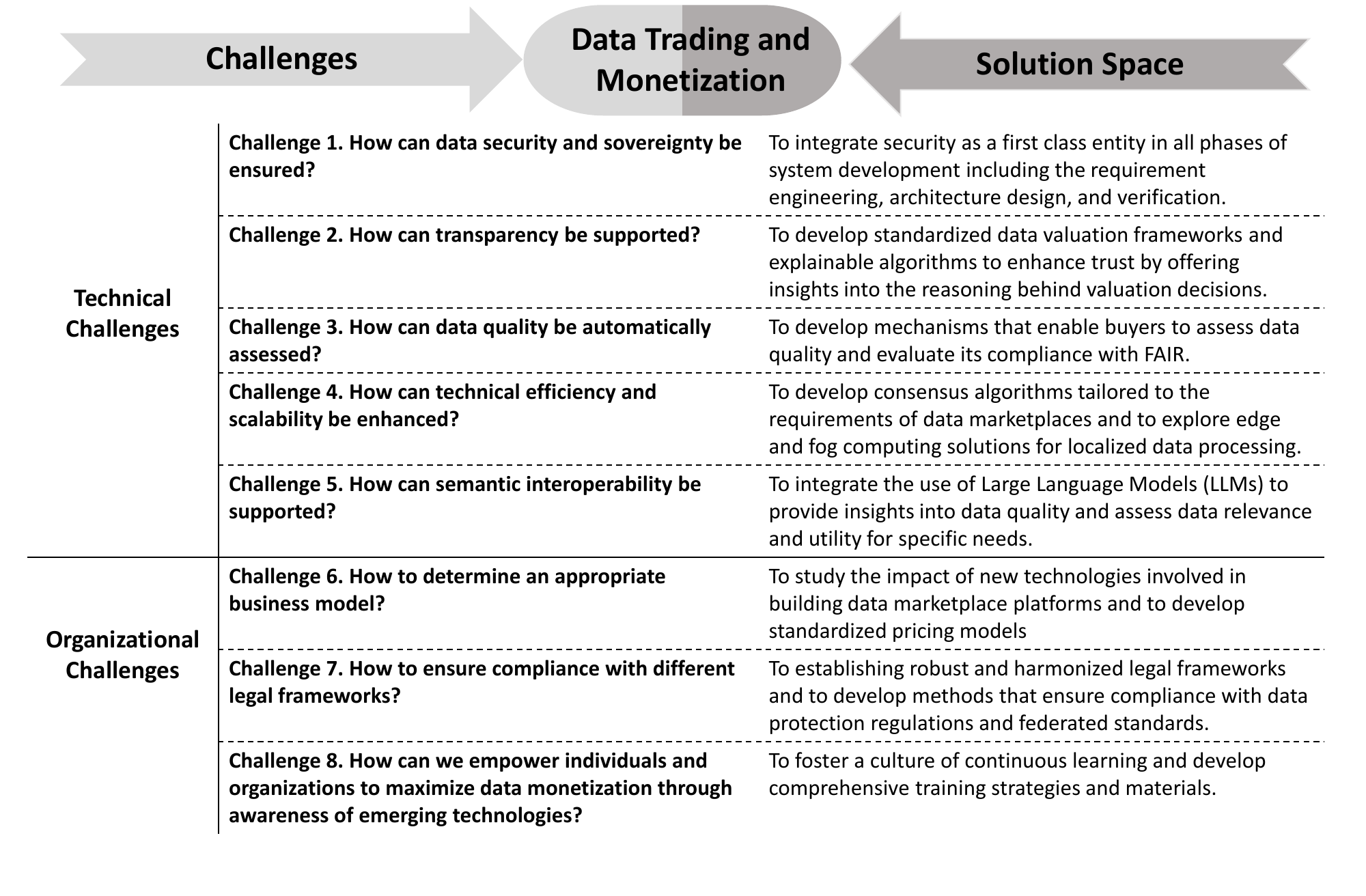}
	\caption{Overview of Challenges and Solutions in the Data Trading and Monetization Landscape.}
	\label{fig:challanges}
\end{figure*}

This section enumerates and engages in a thorough discourse regarding the contemporary challenges that necessitate additional research and development efforts. In order to advance the state of knowledge in the field of data trading and monetization, it is imperative to recognize the areas where current limitations persist and where novel solutions and investigations are required. 

Fig.~\ref{fig:challanges} provides an overview of the identified challenges and solutions in the data trading and monetization landscape. We classify the identified challenges into two categories: \textit{technical challenges}, which pertain to issues related to data infrastructure, security, and analytics, and \textit{organizational challenges}, which encompass hurdles associated with data governance, workforce training, and business models. We have framed the challenges as open research questions, endeavoring to provide prospective solutions for each, in order to facilitate the work of fellow scholars and practitioners in the domain. Detailed descriptions of the technical challenges and organizational challenges are provided in Section~\ref{Sec:tech-challanges} and Section~\ref{Sec:Nontech-challanges}, respectively.

\subsection{Technical Challenges}\label{Sec:tech-challanges}
This section delves into the challenges encountered in the realm of data marketplaces, which are obstacles to data monetization.

\smallskip
\noindent\textbf{Challenge 1. How can data security and sovereignty be ensured?} Similar to any traditional information system that stores and transmits sensitive data, data marketplaces are potential targets for various cybersecurity threats. Therefore, data marketplaces must implement technical and organizational measures to protect data privacy, ensure data integrity, prevent data breaches, address identity and authentication issues, and establish secure data transfer protocols. Additionally, data marketplaces must navigate regulatory compliance requirements, such as data protection laws, to maintain the confidentiality and trustworthiness of the data being exchanged. However, maintaining data security and privacy in data marketplaces is more challenging compared to traditional information systems due to the following factors:

\begin{itemize}
    \item \textit{Data sharing complexity:} Data marketplaces involve the sharing and exchange of data among multiple parties. This complex network of data sharing increases the risk of data exposure, unauthorized access, and breaches during data transmission~\cite{shaabany2016secure, shaabany2018security}.

    \item\textit{Trust and anonymity concerns:} Data marketplaces often require a level of trust and anonymity to encourage participation from various stakeholders. Striking the right balance between maintaining trust and protecting user privacy requires careful design and implementation~\cite{virkar2019investigating}. More specifically, the trade of personal data can result in unintended disclosure of personal information, particularly when individuals act as data providers and large corporations or governments serve as data buyers. This dynamic can create an imbalance of power, increasing the risk of personal information being exposed without the explicit consent or awareness of the individuals involved~\cite{lawrenz2019blockchain,charitsis2018creating}.

    \item\textit{Complexity of controlling data monetization.} The complexity of controlling data monetization poses challenges for data providers in effectively tracking and ensuring compliance with data-sharing agreements. This lack of visibility and control raises concerns among data providers, as they are unable to monitor how their data is being used and whether it aligns with the agreed-upon terms. The fear of competitors benefiting from their data in unforeseen ways further amplifies the apprehension~\cite{spiekermann2019data}. This situation not only creates uncertainties surrounding the intended usage of the data but also introduces potential privacy risks~\cite{koutroumpis2020markets}, as the extent of data utilization and the resulting implications on individual privacy become uncertain.  
\end{itemize}

\begin{myquote}
\noindent\textbf{Toward addressing challenge 1.} Future research should develop comprehensive and dynamic strategies to ensure the consistent safeguarding of data security. Specifically, in the software engineering process of building a data marketplace, it is crucial to give importance to security and privacy through automation support to address several key aspects: Firstly, it involves the elicitation and specification of requirements that effectively capture the essential security and privacy requirements~\cite{shaabany2016secure,abraham2023taxonomy}. Secondly, it requires careful architectural design, ensuring adherence to data protection concerns~\cite{roman2016towards,zichichi2020ensuring}. Thirdly, rigorous validation and verification processes are essential to ensure that the resulting data marketplace meets the specified security and privacy requirements~\cite{ahmadian2018extending}, especially when the data marketplace evolves.
\end{myquote}

\smallskip
\noindent\textbf{Challenge 2. How can transparency be supported?} Ensuring transparency is a significant challenge in data marketplaces, particularly concerning the relationship between data providers and data brokers. There is often a lack of clear and open communication, leading to limited visibility into how data brokers evaluate the value of data and the methodologies employed in this assessment~\cite{oh2019personal}. Data providers often find themselves unaware of the specific criteria or fair processes used by data brokers to determine the worth of their data. This lack of transparency hampers the ability of data providers to make informed decisions about their data and understand the fairness and accuracy of the valuation process~\cite{crain2018limits}. As a result, data providers may feel disadvantaged and uncertain about the fairness and legitimacy of the transactions taking place in the data marketplace ecosystem.

\begin{myquote}
\noindent\textbf{Toward addressing challenge 2.} To overcome the transparency challenge, future research should involve the development of a standardized data valuation framework. The valuation framework should encompass the following components: First, valuation metrics to provide a common language for assessing data and making it easier for both data providers and consumers to comprehend how data is priced~\cite{colli2021translating}. Second, explainable algorithms to enhance trust by offering insights into the reasoning behind valuation decisions, fostering transparency and accountability~\cite{samek2019towards}. Third, blockchain technology for immutable records to ensure the integrity of data transactions~\cite{zheng2017overview}.
\end{myquote}

\smallskip
\noindent\textbf{Challenge 3. How can data quality be assessed in an automated way?} Data marketplaces often struggle with issues related to data accuracy, completeness, and consistency, which are crucial for effective decision-making within data marketplaces~\cite{koutroumpis2020markets}. Additionally, the FAIR principles, advocating for data to be findable, accessible, interoperable, and reusable, present another hurdle~\cite{jacobsen2020fair}. However, one specific challenge lies in the lack of mechanisms that support buyers in assessing data quality or evaluating its adherence to FAIR principles before purchasing. This absence makes it difficult for buyers to determine the accuracy of the data they acquire and compromises their ability to make informed decisions.

\begin{myquote}
\noindent\textbf{Toward addressing challenge 3.} To overcome the challenge of ensuring data quality, efforts are needed to establish mechanisms that enable buyers to assess data quality and evaluate its compliance with FAIR principles before making a purchase~\cite{jacobsen2020fair}. Transparent metadata management systems, standardized data quality assessments, and clear documentation of adherence to FAIR principles can significantly enhance buyer confidence. Implementing such mechanisms would not only facilitate informed decision-making but also promote trust and reliability in market data places, ultimately supporting successful data monetization strategies.
\end{myquote}

\smallskip
\noindent\textbf{Challenge 4. How can technical efficiency and scalability be enhanced?} Data marketplaces utilizing distributed ledger technology or decentralized architectures face increased communication costs due to the constant synchronization and communication required between multiple nodes~\cite{liu2019novel}. The more nodes involved, the higher the communication overhead, which can negatively impact the performance and scalability of the marketplace~\cite{ishmaev2020ethical}. Additionally, the computational complexity of operations such as data verification, encryption, and indexing contributes to substantial computation costs. Meeting these computational demands in a timely and cost-effective manner is essential for ensuring efficiency and scalability in data marketplaces.

\begin{myquote}
\noindent\textbf{Toward addressing challenge 4.} To address the challenge of technical efficiency and scalability, a promising research direction involves optimizing distributed computing. This involves a multifaceted approach, including the investigation of customized consensus algorithms tailored to the unique requirements of data marketplaces such algorithms can enhance data transaction speeds, reduce latency, and ensure data integrity, particularly in large-scale data trading scenarios~\cite{chaudhry2018consensus, mingxiao2017review}. Additionally, exploring edge and fog computing solutions for localized data processing can alleviate the strain on centralized servers~\cite{varghese2020feasibility}, improving overall system efficiency.
\end{myquote}

\smallskip
\noindent\textbf{Challenge 5. How can semantic interoperability be supported?} As previously articulated in~\cite{boukhers2023enhancing}, the cornerstone of effective data exchange is the mutual understanding of the shared data, encapsulated by the concept ``Semantic Interoperability''. However, achieving this understanding is often a hurdle, contingent upon the data consumer's (i.e., data Buyer) grasp of the vocabulary and metadata from the data provider (i.e., data Seller). This necessitates the accessibility of metadata to decode the data content, adhering to the FAIR principles~\cite{wilkinson2016fair}, and the utilization of a common vocabulary. In instances where a common vocabulary is absent, a mapping between the vocabularies of the data consumer and provider is imperative. Illustratively, in a Digital Advertising Dataset, the attribute “\emph{Engagement}” may have different interpretations between the data provider and consumer. For instance, does “\emph{Engagement}” denote a viewer watching an ad video for a duration exceeding a predefined number of seconds? Or, hovering their mouse over an ad for a specific period? Such discrepancies underline the necessity for clear vocabulary mapping to foster effective data exchange and in return data trading and monetization.

\begin{myquote}
\noindent\textbf{Toward addressing challenge 5.} Web technologies have historically tackled the issues of semantic interoperability. However, with the recent surge in machine learning advancements, there are renewed directions to explore and address these challenges more comprehensively~\cite{boukhers2023enhancing}. Notably, the transformative capabilities of transformer architectures, especially Large Language Models (LLMs), uncovered new possibilities. These models can act as a canal, enhancing understanding between data providers and consumers. Such capabilities not only assist in resolving semantic interoperability hurdles but also empower data providers to enhance the trade and monetization potential of their data. Simultaneously, it provides data consumers with insights into data quality, allowing them to assess its relevance and utility for their specific needs.
\end{myquote}

\subsection{Organizational Challenges} \label{Sec:Nontech-challanges}
Beyond the technical challenges discussed in the previous section, it is essential to recognize that substantial organizational obstacles exist for data monetization. Significant non-technical challenges for data monetization are:

\smallskip
\noindent\textbf{Challenge 6. How to determine an appropriate business model?} Organizations recognize the value of their data assets but struggle to determine the most effective and profitable strategies for monetization~\cite{zhang2023survey,spiekermann2019data}. Without established guidelines, pricing strategies, revenue models, and value propositions become ambiguous, hindering the alignment of data monetization efforts with business objectives and leading to missed opportunities for revenue generation. Additionally, the absence of standardized guidelines hampers trust and transparency in data transactions~\cite{spiekermann2019data, calancea2021techniques}. 

However, determining the appropriate price for data products and services is complex due to the intangible nature of data and the absence of standardized pricing models~\cite{vomfell2015classification}. Additionally, there is reluctance among data consumers to pay for data, as organizations struggle to perceive its direct value and return on investment~\cite{niu2020online}. Furthermore, data providers encounter difficulties in formulating an optimal profit strategy that strikes a balance between maximizing revenue and managing the cost structure associated with data acquisition~\cite{mao2019pricing}. This challenge arises from the complexity of assessing the potential returns on data investments and the uncertainty surrounding the monetization potential of their data assets. 

\begin{myquote}
\noindent\textbf{Toward addressing challenge 6.} To address the lack of clear business models, future work should study the impact of new technologies involved in building data marketplace platforms such as blockchain and IDS connectors on the business models. This will enable the development of well-defined business model guidelines to foster trust, ensure fair transactions, and provide clarity and transparency in data monetization, enabling organizations to maximize the value of their data assets~\cite{calancea2021techniques,van2021creating}. In addition, bridging the gap in price acceptance requires efforts to educate about the value of data, raise awareness of its potential impact, and develop standardized pricing models~\cite{mao2019pricing,niu2020online,chen2019demonstration}.
\end{myquote}

\smallskip
\noindent\textbf{Challenge 7. How to ensure compliance with different legal frameworks?} Data is subject to various legal considerations, including privacy, security, intellectual property, and data protection laws. However, the evolving nature of data and rapid technological advancements often outpace the development of comprehensive legal guidelines~\cite{spiekermann2019data}. This leads to ambiguity and uncertainty regarding data rights, ownership, and consent~\cite{sorlie2019sensing}. For example, data ownership and consent become more intricate in data marketplaces where data is sourced from multiple providers. Specifically, some scholars suggest that data (especially personal data) should be owned by individuals. However, alternative viewpoints from legal scholars argue that data ownership may be unfeasible or conceptually flawed in addressing the complex societal and economic challenges presented by the data economy~\cite{evans2011much}. Therefore, the absence of clear legal frameworks hinders organizations from leveraging data and raises risks of legal disputes and compliance issues. 

\begin{myquote}
\noindent\textbf{Toward addressing challenge 7.} To address the challenge of the lack of legal frameworks, future research should concentrate on:  First, Establishing robust and harmonized legal frameworks to provide clarity, protect rights, ensure privacy, and facilitate fair and secure data monetization practices. Second, developing methods to ensure compliance or (at least) alignment with (i) data protection regulations such as GDPR to protect individuals' privacy, (ii) federated standards for data infrastructure such as Gaia-X to promote interoperability and data sovereignty, and (iii) reference architecture models such as the IDS-RAM to establish standardized data management practices. 
\end{myquote}

\smallskip
\noindent\textbf{Challenge 8. How can we empower individuals and organizations to maximize data monetization through awareness of emerging technologies?} The rapid advancements in technology have introduced innovative tools and platforms that can enhance data monetization efforts, but the shortage of accessible and up-to-date training resources limits individuals' ability to acquire the knowledge and competencies required for success. This skills gap hampers organizations' capacity to fully leverage data monetization and capitalize on the opportunities presented by emerging technologies. Addressing this challenge is crucial to bridge the skills gap and empower workers in the field of data monetization. 

\begin{myquote}
\noindent\textbf{Toward addressing challenge 8.} To address the skills gap organizations and companies should focus on developing comprehensive training strategies and materials, so organizations can provide individuals with the necessary guidance and expertise to navigate and leverage emerging technologies within data marketplaces effectively. Additionally, fostering a culture of continuous learning and providing ongoing training opportunities can help workers stay updated with the latest advancements and adapt to evolving technologies.
\end{myquote}


\section{Research Limitations}\label{sec:discussion}
The findings of our research are subject to several threats.  A threat to \textit{external validity} is that we discuss the current challenges in the field of data trading and monetization solely from academic perspectives, without the direct involvement of industrial experts. Although our knowledge closely aligns with the experiences of industrial experts, this approach introduces a potential limitation in terms of the practical applicability of our suggestions for addressing the identified challenges in real-world settings. To mitigate this concern, future work should involve consultation with industry experts and compare their insights with our findings.

A threat to \textit{internal validity} is that we have limited the scope of the state-of-the-art to recent developments in the field of data trading within the European Union (EU). However, this focused approach poses challenges in terms of generalizability. For example, ensuring compliance with specific standards within the EU, such as Gaia-X and IDS, may not be a requirement for data trading practices in other regions or on a global scale. As a consequence, the generalizability of our findings to broader contexts may be compromised. To address this limitation, a systematic literature review is required to investigate data trading developments in various regions, facilitating a more comprehensive understanding of the challenges within this dynamic field.

\section{Conclusions} 
In this paper, we have examined the landscape of data monetization, highlighting both technical and non-technical challenges that data marketplaces face. The rapid growth in data generation and technological advancements has opened up opportunities for data monetization, but it has also introduced complex hurdles that need to be addressed for successful implementation.

On the technical side, challenges related to data security, privacy, transparency, data quality, and technical efficiency have been discussed. These challenges underscore the need for robust technological solutions and standards to ensure the integrity and effectiveness of data marketplaces. Addressing these issues will be crucial in building trust among data providers and consumers.

On the organizational side, challenges such as the lack of clear business models and legal frameworks, have also been identified as significant barriers to data monetization. The lack of attention given to these organizational challenges hinders the full realization of the potential of data monetization. Therefore, it is imperative to expand the focus of research and development efforts to encompass these organizational challenges and devise comprehensive strategies that ensure the successful implementation and sustainable growth of data monetization initiatives.\label{sec:conclusion}


\section*{Acknowledgment}
This research was partially supported by the \textit{EU’s Horizon Digital, Industry, and Space program} under grant agreement ID 101092989 -DATAMITE.

The contributions of \emph{Zeyd Boukhers} and \emph{Christoph Lange} to this work were funded by the \emph{FAIR Data Spaces} project of the German Federal Ministry of Education and Research (BMBF) under grant numbers \textbf{FAIRDS05} and \textbf{FAIRDS15}.\label{sec:Acknowledgment}


\bibliographystyle{ACM-Reference-Format}
\bibliography{sample-base}


\begin{thebibliography}{51}


\ifx \showCODEN    \undefined \def \showCODEN     #1{\unskip}     \fi
\ifx \showDOI      \undefined \def \showDOI       #1{#1}\fi
\ifx \showISBNx    \undefined \def \showISBNx     #1{\unskip}     \fi
\ifx \showISBNxiii \undefined \def \showISBNxiii  #1{\unskip}     \fi
\ifx \showISSN     \undefined \def \showISSN      #1{\unskip}     \fi
\ifx \showLCCN     \undefined \def \showLCCN      #1{\unskip}     \fi
\ifx \shownote     \undefined \def \shownote      #1{#1}          \fi
\ifx \showarticletitle \undefined \def \showarticletitle #1{#1}   \fi
\ifx \showURL      \undefined \def \showURL       {\relax}        \fi
\providecommand\bibfield[2]{#2}
\providecommand\bibinfo[2]{#2}
\providecommand\natexlab[1]{#1}
\providecommand\showeprint[2][]{arXiv:#2}

\bibitem[Abraham et~al\mbox{.}(2023)]%
        {abraham2023taxonomy}
\bibfield{author}{\bibinfo{person}{Rene Abraham}, \bibinfo{person}{Johannes Schneider}, {and} \bibinfo{person}{Jan vom Brocke}.} \bibinfo{year}{2023}\natexlab{}.
\newblock \showarticletitle{A taxonomy of data governance decision domains in data marketplaces}.
\newblock \bibinfo{journal}{\emph{Electronic Markets}} \bibinfo{volume}{33}, \bibinfo{number}{1} (\bibinfo{year}{2023}), \bibinfo{pages}{22}.
\newblock


\bibitem[Ahmadian et~al\mbox{.}(2018)]%
        {ahmadian2018extending}
\bibfield{author}{\bibinfo{person}{Amir~Shayan Ahmadian}, \bibinfo{person}{Jan J{\"u}rjens}, {and} \bibinfo{person}{Daniel Str{\"u}ber}.} \bibinfo{year}{2018}\natexlab{}.
\newblock \showarticletitle{Extending model-based privacy analysis for the industrial data space by exploiting privacy level agreements}. In \bibinfo{booktitle}{\emph{Proceedings of the 33rd annual ACM symposium on applied computing}}. \bibinfo{pages}{1142--1149}.
\newblock


\bibitem[Bhujel and Rahulamathavan(2022)]%
        {bhujel2022survey}
\bibfield{author}{\bibinfo{person}{Sangam Bhujel} {and} \bibinfo{person}{Yogachandran Rahulamathavan}.} \bibinfo{year}{2022}\natexlab{}.
\newblock \showarticletitle{A survey: Security, transparency, and scalability issues of nft’s and its marketplaces}.
\newblock \bibinfo{journal}{\emph{Sensors}} \bibinfo{volume}{22}, \bibinfo{number}{22} (\bibinfo{year}{2022}), \bibinfo{pages}{8833}.
\newblock


\bibitem[Boukhers et~al\mbox{.}(2023)]%
        {boukhers2023enhancing}
\bibfield{author}{\bibinfo{person}{Zeyd Boukhers}, \bibinfo{person}{Christoph Lange}, {and} \bibinfo{person}{Oya Beyan}.} \bibinfo{year}{2023}\natexlab{}.
\newblock \showarticletitle{Enhancing Data Space Semantic Interoperability through Machine Learning: a Visionary Perspective}. In \bibinfo{booktitle}{\emph{Companion Proceedings of the ACM Web Conference 2023}}. \bibinfo{pages}{1462--1467}.
\newblock


\bibitem[Braud et~al\mbox{.}(2021)]%
        {braud2021road}
\bibfield{author}{\bibinfo{person}{Arnaud Braud}, \bibinfo{person}{Ga{\"e}l Fromentoux}, \bibinfo{person}{Benoit Radier}, {and} \bibinfo{person}{Olivier Le~Grand}.} \bibinfo{year}{2021}\natexlab{}.
\newblock \showarticletitle{The road to European digital sovereignty with Gaia-X and IDSA}.
\newblock \bibinfo{journal}{\emph{IEEE network}} \bibinfo{volume}{35}, \bibinfo{number}{2} (\bibinfo{year}{2021}), \bibinfo{pages}{4--5}.
\newblock


\bibitem[Calancea and Alboaie(2021)]%
        {calancea2021techniques}
\bibfield{author}{\bibinfo{person}{Cristina~Georgiana Calancea} {and} \bibinfo{person}{Lenuța Alboaie}.} \bibinfo{year}{2021}\natexlab{}.
\newblock \showarticletitle{Techniques to improve b2b data governance using fair principles}.
\newblock \bibinfo{journal}{\emph{Mathematics}} \bibinfo{volume}{9}, \bibinfo{number}{9} (\bibinfo{year}{2021}), \bibinfo{pages}{1059}.
\newblock


\bibitem[Catalini and Gans(2020)]%
        {catalini2020some}
\bibfield{author}{\bibinfo{person}{Christian Catalini} {and} \bibinfo{person}{Joshua~S Gans}.} \bibinfo{year}{2020}\natexlab{}.
\newblock \showarticletitle{Some simple economics of the blockchain}.
\newblock \bibinfo{journal}{\emph{Commun. ACM}} \bibinfo{volume}{63}, \bibinfo{number}{7} (\bibinfo{year}{2020}), \bibinfo{pages}{80--90}.
\newblock


\bibitem[Charitsis et~al\mbox{.}(2018)]%
        {charitsis2018creating}
\bibfield{author}{\bibinfo{person}{Vassilis Charitsis}, \bibinfo{person}{Detlev Zwick}, {and} \bibinfo{person}{Alan Bradshaw}.} \bibinfo{year}{2018}\natexlab{}.
\newblock \showarticletitle{Creating worlds that create audiences: Theorising personal data markets in the age of communicative capitalism}.
\newblock  (\bibinfo{year}{2018}).
\newblock


\bibitem[Chaudhry and Yousaf(2018)]%
        {chaudhry2018consensus}
\bibfield{author}{\bibinfo{person}{Natalia Chaudhry} {and} \bibinfo{person}{Muhammad~Murtaza Yousaf}.} \bibinfo{year}{2018}\natexlab{}.
\newblock \showarticletitle{Consensus algorithms in blockchain: Comparative analysis, challenges and opportunities}. In \bibinfo{booktitle}{\emph{2018 12th International Conference on Open Source Systems and Technologies (ICOSST)}}. IEEE, \bibinfo{pages}{54--63}.
\newblock


\bibitem[Chen et~al\mbox{.}(2019)]%
        {chen2019demonstration}
\bibfield{author}{\bibinfo{person}{Lingjiao Chen}, \bibinfo{person}{Hongyi Wang}, \bibinfo{person}{Leshang Chen}, \bibinfo{person}{Paraschos Koutris}, {and} \bibinfo{person}{Arun Kumar}.} \bibinfo{year}{2019}\natexlab{}.
\newblock \showarticletitle{Demonstration of Nimbus: Model-based Pricing for Machine Learning in a Data Marketplace}. In \bibinfo{booktitle}{\emph{Proceedings of the 2019 International Conference on Management of Data}}. \bibinfo{pages}{1885--1888}.
\newblock


\bibitem[Colli et~al\mbox{.}(2021)]%
        {colli2021translating}
\bibfield{author}{\bibinfo{person}{Michele Colli}, \bibinfo{person}{Jonas Nygaard~Uhrenholt}, \bibinfo{person}{Ole Madsen}, {and} \bibinfo{person}{Brian~Vejrum Waehrens}.} \bibinfo{year}{2021}\natexlab{}.
\newblock \showarticletitle{Translating transparency into value: an approach to design IoT solutions}.
\newblock \bibinfo{journal}{\emph{Journal of Manufacturing Technology Management}} \bibinfo{volume}{32}, \bibinfo{number}{8} (\bibinfo{year}{2021}), \bibinfo{pages}{1515--1532}.
\newblock


\bibitem[Crain(2018)]%
        {crain2018limits}
\bibfield{author}{\bibinfo{person}{Matthew Crain}.} \bibinfo{year}{2018}\natexlab{}.
\newblock \showarticletitle{The limits of transparency: Data brokers and commodification}.
\newblock \bibinfo{journal}{\emph{new media \& society}} \bibinfo{volume}{20}, \bibinfo{number}{1} (\bibinfo{year}{2018}), \bibinfo{pages}{88--104}.
\newblock


\bibitem[Evans(2011)]%
        {evans2011much}
\bibfield{author}{\bibinfo{person}{Barbara~J Evans}.} \bibinfo{year}{2011}\natexlab{}.
\newblock \showarticletitle{Much ado about data ownership}.
\newblock \bibinfo{journal}{\emph{Harv. JL \& Tech.}}  \bibinfo{volume}{25} (\bibinfo{year}{2011}), \bibinfo{pages}{69}.
\newblock


\bibitem[Evans(2014)]%
        {evans2014economic}
\bibfield{author}{\bibinfo{person}{David~S Evans}.} \bibinfo{year}{2014}\natexlab{}.
\newblock \showarticletitle{Economic aspects of Bitcoin and other decentralized public-ledger currency platforms}.
\newblock \bibinfo{journal}{\emph{University of Chicago Coase-Sandor Institute for Law \& Economics Research Paper}} \bibinfo{number}{685} (\bibinfo{year}{2014}).
\newblock


\bibitem[Hynes et~al\mbox{.}(2018)]%
        {hynes2018demonstration}
\bibfield{author}{\bibinfo{person}{Nick Hynes}, \bibinfo{person}{David Dao}, \bibinfo{person}{David Yan}, \bibinfo{person}{Raymond Cheng}, {and} \bibinfo{person}{Dawn Song}.} \bibinfo{year}{2018}\natexlab{}.
\newblock \showarticletitle{A demonstration of sterling: a privacy-preserving data marketplace}.
\newblock \bibinfo{journal}{\emph{Proceedings of the VLDB Endowment}} \bibinfo{volume}{11}, \bibinfo{number}{12} (\bibinfo{year}{2018}), \bibinfo{pages}{2086--2089}.
\newblock


\bibitem[Ishmaev(2020)]%
        {ishmaev2020ethical}
\bibfield{author}{\bibinfo{person}{Georgy Ishmaev}.} \bibinfo{year}{2020}\natexlab{}.
\newblock \showarticletitle{The ethical limits of blockchain-enabled markets for private IoT data}.
\newblock \bibinfo{journal}{\emph{Philosophy \& Technology}}  \bibinfo{volume}{33} (\bibinfo{year}{2020}), \bibinfo{pages}{411--432}.
\newblock


\bibitem[Jacobsen et~al\mbox{.}(2020)]%
        {jacobsen2020fair}
\bibfield{author}{\bibinfo{person}{Annika Jacobsen}, \bibinfo{person}{Ricardo de Miranda~Azevedo}, \bibinfo{person}{Nick Juty}, \bibinfo{person}{Dominique Batista}, \bibinfo{person}{Simon Coles}, \bibinfo{person}{Ronald Cornet}, \bibinfo{person}{M{\'e}lanie Courtot}, \bibinfo{person}{Merc{\`e} Crosas}, \bibinfo{person}{Michel Dumontier}, \bibinfo{person}{Chris~T Evelo}, {et~al\mbox{.}}} \bibinfo{year}{2020}\natexlab{}.
\newblock \bibinfo{title}{FAIR principles: interpretations and implementation considerations}.
\newblock , \bibinfo{numpages}{10--29}~pages.
\newblock


\bibitem[Koutroumpis et~al\mbox{.}(2017)]%
        {koutroumpis2017unfulfilled}
\bibfield{author}{\bibinfo{person}{Pantelis Koutroumpis}, \bibinfo{person}{Aija Leiponen}, {and} \bibinfo{person}{Llewellyn~DW Thomas}.} \bibinfo{year}{2017}\natexlab{}.
\newblock \bibinfo{booktitle}{\emph{The (unfulfilled) potential of data marketplaces}}.
\newblock \bibinfo{type}{{T}echnical {R}eport}. \bibinfo{institution}{ETLA working papers}.
\newblock


\bibitem[Koutroumpis et~al\mbox{.}(2020)]%
        {koutroumpis2020markets}
\bibfield{author}{\bibinfo{person}{Pantelis Koutroumpis}, \bibinfo{person}{Aija Leiponen}, {and} \bibinfo{person}{Llewellyn~DW Thomas}.} \bibinfo{year}{2020}\natexlab{}.
\newblock \showarticletitle{Markets for data}.
\newblock \bibinfo{journal}{\emph{Industrial and Corporate Change}} \bibinfo{volume}{29}, \bibinfo{number}{3} (\bibinfo{year}{2020}), \bibinfo{pages}{645--660}.
\newblock


\bibitem[Lawrenz et~al\mbox{.}(2019)]%
        {lawrenz2019blockchain}
\bibfield{author}{\bibinfo{person}{Sebastian Lawrenz}, \bibinfo{person}{Priyanka Sharma}, {and} \bibinfo{person}{Andreas Rausch}.} \bibinfo{year}{2019}\natexlab{}.
\newblock \showarticletitle{Blockchain technology as an approach for data marketplaces}. In \bibinfo{booktitle}{\emph{Proceedings of the 2019 international conference on blockchain technology}}. \bibinfo{pages}{55--59}.
\newblock


\bibitem[Li et~al\mbox{.}(2020)]%
        {li2020blockchain}
\bibfield{author}{\bibinfo{person}{Yuzheng Li}, \bibinfo{person}{Chuan Chen}, \bibinfo{person}{Nan Liu}, \bibinfo{person}{Huawei Huang}, \bibinfo{person}{Zibin Zheng}, {and} \bibinfo{person}{Qiang Yan}.} \bibinfo{year}{2020}\natexlab{}.
\newblock \showarticletitle{A blockchain-based decentralized federated learning framework with committee consensus}.
\newblock \bibinfo{journal}{\emph{IEEE Network}} \bibinfo{volume}{35}, \bibinfo{number}{1} (\bibinfo{year}{2020}), \bibinfo{pages}{234--241}.
\newblock


\bibitem[Liu et~al\mbox{.}(2019)]%
        {liu2019novel}
\bibfield{author}{\bibinfo{person}{Kang Liu}, \bibinfo{person}{Wuhui Chen}, \bibinfo{person}{Zibin Zheng}, \bibinfo{person}{Zhenni Li}, {and} \bibinfo{person}{Wei Liang}.} \bibinfo{year}{2019}\natexlab{}.
\newblock \showarticletitle{A novel debt-credit mechanism for blockchain-based data-trading in internet of vehicles}.
\newblock \bibinfo{journal}{\emph{IEEE Internet of Things Journal}} \bibinfo{volume}{6}, \bibinfo{number}{5} (\bibinfo{year}{2019}), \bibinfo{pages}{9098--9111}.
\newblock


\bibitem[Mao et~al\mbox{.}(2019)]%
        {mao2019pricing}
\bibfield{author}{\bibinfo{person}{Weichao Mao}, \bibinfo{person}{Zhenzhe Zheng}, {and} \bibinfo{person}{Fan Wu}.} \bibinfo{year}{2019}\natexlab{}.
\newblock \showarticletitle{Pricing for revenue maximization in iot data markets: An information design perspective}. In \bibinfo{booktitle}{\emph{IEEE INFOCOM 2019-IEEE Conference on Computer Communications}}. IEEE, \bibinfo{pages}{1837--1845}.
\newblock


\bibitem[Mingxiao et~al\mbox{.}(2017)]%
        {mingxiao2017review}
\bibfield{author}{\bibinfo{person}{Du Mingxiao}, \bibinfo{person}{Ma Xiaofeng}, \bibinfo{person}{Zhang Zhe}, \bibinfo{person}{Wang Xiangwei}, {and} \bibinfo{person}{Chen Qijun}.} \bibinfo{year}{2017}\natexlab{}.
\newblock \showarticletitle{A review on consensus algorithm of blockchain}. In \bibinfo{booktitle}{\emph{2017 IEEE international conference on systems, man, and cybernetics (SMC)}}. IEEE, \bibinfo{pages}{2567--2572}.
\newblock


\bibitem[Niebel et~al\mbox{.}(2022)]%
        {niebel2022gaia}
\bibfield{author}{\bibinfo{person}{Crispin Niebel}, \bibinfo{person}{Abel Reiberg}, {and} \bibinfo{person}{Peter Kraemer}.} \bibinfo{year}{2022}\natexlab{}.
\newblock \bibinfo{booktitle}{\emph{Gaia-X for SMEs}}.
\newblock \bibinfo{type}{White Paper}. \bibinfo{institution}{Gaia-X Hub Germany}.
\newblock


\bibitem[Niu et~al\mbox{.}(2020)]%
        {niu2020online}
\bibfield{author}{\bibinfo{person}{Chaoyue Niu}, \bibinfo{person}{Zhenzhe Zheng}, \bibinfo{person}{Fan Wu}, \bibinfo{person}{Shaojie Tang}, {and} \bibinfo{person}{Guihai Chen}.} \bibinfo{year}{2020}\natexlab{}.
\newblock \showarticletitle{Online pricing with reserve price constraint for personal data markets}.
\newblock \bibinfo{journal}{\emph{IEEE Transactions on Knowledge and Data Engineering}} \bibinfo{volume}{34}, \bibinfo{number}{4} (\bibinfo{year}{2020}), \bibinfo{pages}{1928--1943}.
\newblock


\bibitem[Oh et~al\mbox{.}(2019)]%
        {oh2019personal}
\bibfield{author}{\bibinfo{person}{Hyeontaek Oh}, \bibinfo{person}{Sangdon Park}, \bibinfo{person}{Gyu~Myoung Lee}, \bibinfo{person}{Hwanjo Heo}, {and} \bibinfo{person}{Jun~Kyun Choi}.} \bibinfo{year}{2019}\natexlab{}.
\newblock \showarticletitle{Personal data trading scheme for data brokers in IoT data marketplaces}.
\newblock \bibinfo{journal}{\emph{IEEE Access}}  \bibinfo{volume}{7} (\bibinfo{year}{2019}), \bibinfo{pages}{40120--40132}.
\newblock


\bibitem[Otto et~al\mbox{.}(2019)]%
        {otto2019international}
\bibfield{author}{\bibinfo{person}{Boris Otto}, \bibinfo{person}{Michael ten Hompel}, {and} \bibinfo{person}{Stefan Wrobel}.} \bibinfo{year}{2019}\natexlab{}.
\newblock \showarticletitle{International Data Spaces: Reference architecture for the digitization of industries}.
\newblock \bibinfo{journal}{\emph{Digital transformation}} (\bibinfo{year}{2019}), \bibinfo{pages}{109--128}.
\newblock


\bibitem[Pettenpohl et~al\mbox{.}(2022)]%
        {pettenpohl2022international}
\bibfield{author}{\bibinfo{person}{Heinrich Pettenpohl}, \bibinfo{person}{Markus Spiekermann}, {and} \bibinfo{person}{Jan~Ruben Both}.} \bibinfo{year}{2022}\natexlab{}.
\newblock \showarticletitle{International data spaces in a nutshell}.
\newblock \bibinfo{journal}{\emph{Designing Data Spaces; Springer: Cham, Switzerland}} (\bibinfo{year}{2022}), \bibinfo{pages}{29--40}.
\newblock


\bibitem[Ramachandran et~al\mbox{.}(2018)]%
        {ramachandran2018towards}
\bibfield{author}{\bibinfo{person}{Gowri~Sankar Ramachandran}, \bibinfo{person}{Rahul Radhakrishnan}, {and} \bibinfo{person}{Bhaskar Krishnamachari}.} \bibinfo{year}{2018}\natexlab{}.
\newblock \showarticletitle{Towards a decentralized data marketplace for smart cities}. In \bibinfo{booktitle}{\emph{2018 IEEE International Smart Cities Conference (ISC2)}}. IEEE, \bibinfo{pages}{1--8}.
\newblock


\bibitem[Roman and Stefano(2016)]%
        {roman2016towards}
\bibfield{author}{\bibinfo{person}{Dumitru Roman} {and} \bibinfo{person}{Gatti Stefano}.} \bibinfo{year}{2016}\natexlab{}.
\newblock \showarticletitle{Towards a reference architecture for trusted data marketplaces: The credit scoring perspective}. In \bibinfo{booktitle}{\emph{2016 2nd International Conference on Open and Big Data (OBD)}}. IEEE, \bibinfo{pages}{95--101}.
\newblock


\bibitem[Rydning et~al\mbox{.}(2018)]%
        {rydning2018digitization}
\bibfield{author}{\bibinfo{person}{David Reinsel-John Gantz-John Rydning}, \bibinfo{person}{John Reinsel}, {and} \bibinfo{person}{John Gantz}.} \bibinfo{year}{2018}\natexlab{}.
\newblock \showarticletitle{The digitization of the world from edge to core}.
\newblock \bibinfo{journal}{\emph{Framingham: International Data Corporation}}  \bibinfo{volume}{16} (\bibinfo{year}{2018}), \bibinfo{pages}{1--28}.
\newblock


\bibitem[Samek and M{\"u}ller(2019)]%
        {samek2019towards}
\bibfield{author}{\bibinfo{person}{Wojciech Samek} {and} \bibinfo{person}{Klaus-Robert M{\"u}ller}.} \bibinfo{year}{2019}\natexlab{}.
\newblock \showarticletitle{Towards explainable artificial intelligence}.
\newblock \bibinfo{journal}{\emph{Explainable AI: interpreting, explaining and visualizing deep learning}} (\bibinfo{year}{2019}), \bibinfo{pages}{5--22}.
\newblock


\bibitem[Shaabany and Anderl(2018)]%
        {shaabany2018security}
\bibfield{author}{\bibinfo{person}{Ghaidaa Shaabany} {and} \bibinfo{person}{Reiner Anderl}.} \bibinfo{year}{2018}\natexlab{}.
\newblock \showarticletitle{Security by design as an approach to design a secure industry 4.0-capable machine enabling online-trading of technology data}. In \bibinfo{booktitle}{\emph{2018 International Conference on System Science and Engineering (ICSSE)}}. IEEE, \bibinfo{pages}{1--5}.
\newblock


\bibitem[Shaabany et~al\mbox{.}(2016)]%
        {shaabany2016secure}
\bibfield{author}{\bibinfo{person}{Ghaidaa Shaabany}, \bibinfo{person}{Marco Grimm}, {and} \bibinfo{person}{Reiner Anderl}.} \bibinfo{year}{2016}\natexlab{}.
\newblock \showarticletitle{Secure information model for data marketplaces enabling global distributed manufacturing}.
\newblock \bibinfo{journal}{\emph{Procedia CIRP}}  \bibinfo{volume}{50} (\bibinfo{year}{2016}), \bibinfo{pages}{360--365}.
\newblock


\bibitem[S{\o}rlie and Altmann(2019)]%
        {sorlie2019sensing}
\bibfield{author}{\bibinfo{person}{Jan-Terje S{\o}rlie} {and} \bibinfo{person}{J{\"o}rn Altmann}.} \bibinfo{year}{2019}\natexlab{}.
\newblock \showarticletitle{Sensing as a service revisited: A property rights enforcement and pricing model for IIoT data marketplaces}. In \bibinfo{booktitle}{\emph{Economics of Grids, Clouds, Systems, and Services: 16th International Conference, GECON 2019, Leeds, UK, September 17--19, 2019, Proceedings 16}}. Springer, \bibinfo{pages}{127--139}.
\newblock


\bibitem[Spiekermann(2019)]%
        {spiekermann2019data}
\bibfield{author}{\bibinfo{person}{Markus Spiekermann}.} \bibinfo{year}{2019}\natexlab{}.
\newblock \showarticletitle{Data marketplaces: Trends and monetisation of data goods}.
\newblock \bibinfo{journal}{\emph{Intereconomics}} \bibinfo{volume}{54}, \bibinfo{number}{4} (\bibinfo{year}{2019}), \bibinfo{pages}{208--216}.
\newblock


\bibitem[Swan(2015)]%
        {swan2015blockchain}
\bibfield{author}{\bibinfo{person}{Melanie Swan}.} \bibinfo{year}{2015}\natexlab{}.
\newblock \bibinfo{booktitle}{\emph{Blockchain: Blueprint for a new economy}}.
\newblock \bibinfo{publisher}{" O'Reilly Media, Inc."}.
\newblock


\bibitem[Tardieu(2022)]%
        {tardieu2022role}
\bibfield{author}{\bibinfo{person}{Hubert Tardieu}.} \bibinfo{year}{2022}\natexlab{}.
\newblock \showarticletitle{Role of Gaia-X in the European Data Space Ecosystem}.
\newblock In \bibinfo{booktitle}{\emph{Designing Data Spaces: The Ecosystem Approach To Competitive Advantage}}. \bibinfo{publisher}{Springer International Publishing Cham}, \bibinfo{pages}{41--59}.
\newblock


\bibitem[Thomas and Leiponen(2016)]%
        {thomas2016big}
\bibfield{author}{\bibinfo{person}{Llewellyn~DW Thomas} {and} \bibinfo{person}{Aija Leiponen}.} \bibinfo{year}{2016}\natexlab{}.
\newblock \showarticletitle{Big data commercialization}.
\newblock \bibinfo{journal}{\emph{IEEE Engineering Management Review}} \bibinfo{volume}{44}, \bibinfo{number}{2} (\bibinfo{year}{2016}), \bibinfo{pages}{74--90}.
\newblock


\bibitem[Upadhyay et~al\mbox{.}(2021)]%
        {upadhyay2021blockchain}
\bibfield{author}{\bibinfo{person}{Arvind Upadhyay}, \bibinfo{person}{Sumona Mukhuty}, \bibinfo{person}{Vikas Kumar}, {and} \bibinfo{person}{Yigit Kazancoglu}.} \bibinfo{year}{2021}\natexlab{}.
\newblock \showarticletitle{Blockchain technology and the circular economy: Implications for sustainability and social responsibility}.
\newblock \bibinfo{journal}{\emph{Journal of cleaner production}}  \bibinfo{volume}{293} (\bibinfo{year}{2021}), \bibinfo{pages}{126130}.
\newblock


\bibitem[van~de Ven et~al\mbox{.}(2021)]%
        {van2021creating}
\bibfield{author}{\bibinfo{person}{Montijn van~de Ven}, \bibinfo{person}{Antragama~Ewa Abbas}, \bibinfo{person}{Zenlin Kwee}, {and} \bibinfo{person}{Mark de Reuver}.} \bibinfo{year}{2021}\natexlab{}.
\newblock \showarticletitle{Creating a taxonomy of business models for data marketplaces}. In \bibinfo{booktitle}{\emph{34th Bled EConference: Digital Support from Crisis to Progressive Change}}. University of Maribor Press, \bibinfo{pages}{313--325}.
\newblock


\bibitem[Varghese et~al\mbox{.}(2020)]%
        {varghese2020feasibility}
\bibfield{author}{\bibinfo{person}{Blesson Varghese}, \bibinfo{person}{Nan Wang}, \bibinfo{person}{Dimitrios~S Nikolopoulos}, {and} \bibinfo{person}{Rajkumar Buyya}.} \bibinfo{year}{2020}\natexlab{}.
\newblock \showarticletitle{Feasibility of fog computing}.
\newblock \bibinfo{journal}{\emph{Handbook of Integration of Cloud Computing, Cyber Physical Systems and Internet of Things}} (\bibinfo{year}{2020}), \bibinfo{pages}{127--146}.
\newblock


\bibitem[Virkar et~al\mbox{.}(2019)]%
        {virkar2019investigating}
\bibfield{author}{\bibinfo{person}{Shefali Virkar}, \bibinfo{person}{Gabriela Viale~Pereira}, {and} \bibinfo{person}{Michela Vignoli}.} \bibinfo{year}{2019}\natexlab{}.
\newblock \showarticletitle{Investigating the social, political, economic and cultural implications of data trading}. In \bibinfo{booktitle}{\emph{Electronic Government: 18th IFIP WG 8.5 International Conference, EGOV 2019, San Benedetto Del Tronto, Italy, September 2--4, 2019, Proceedings 18}}. Springer, \bibinfo{pages}{215--229}.
\newblock


\bibitem[Vomfell et~al\mbox{.}(2015)]%
        {vomfell2015classification}
\bibfield{author}{\bibinfo{person}{Lara Vomfell}, \bibinfo{person}{Florian Stahl}, \bibinfo{person}{Fabian Schomm}, {and} \bibinfo{person}{Gottfried Vossen}.} \bibinfo{year}{2015}\natexlab{}.
\newblock \bibinfo{booktitle}{\emph{A classification framework for data marketplaces}}.
\newblock \bibinfo{type}{{T}echnical {R}eport}. \bibinfo{institution}{Ercis working paper}.
\newblock


\bibitem[Wilkinson et~al\mbox{.}(2016)]%
        {wilkinson2016fair}
\bibfield{author}{\bibinfo{person}{Mark~D Wilkinson}, \bibinfo{person}{Michel Dumontier}, \bibinfo{person}{IJsbrand~Jan Aalbersberg}, \bibinfo{person}{Gabrielle Appleton}, \bibinfo{person}{Myles Axton}, \bibinfo{person}{Arie Baak}, \bibinfo{person}{Niklas Blomberg}, \bibinfo{person}{Jan-Willem Boiten}, \bibinfo{person}{Luiz~Bonino da Silva~Santos}, \bibinfo{person}{Philip~E Bourne}, {et~al\mbox{.}}} \bibinfo{year}{2016}\natexlab{}.
\newblock \showarticletitle{The FAIR Guiding Principles for scientific data management and stewardship}.
\newblock \bibinfo{journal}{\emph{Scientific data}} \bibinfo{volume}{3}, \bibinfo{number}{1} (\bibinfo{year}{2016}), \bibinfo{pages}{1--9}.
\newblock


\bibitem[Xu and Chen(2021)]%
        {xu2021fed}
\bibfield{author}{\bibinfo{person}{Ronghua Xu} {and} \bibinfo{person}{Yu Chen}.} \bibinfo{year}{2021}\natexlab{}.
\newblock \showarticletitle{Fed-ddm: A federated ledgers based framework for hierarchical decentralized data marketplaces}. In \bibinfo{booktitle}{\emph{2021 international conference on computer communications and networks (ICCCN)}}. IEEE, \bibinfo{pages}{1--8}.
\newblock


\bibitem[Zhang et~al\mbox{.}(2023)]%
        {zhang2023survey}
\bibfield{author}{\bibinfo{person}{Mengxiao Zhang}, \bibinfo{person}{Fernando Beltr{\'a}n}, {and} \bibinfo{person}{Jiamou Liu}.} \bibinfo{year}{2023}\natexlab{}.
\newblock \showarticletitle{A Survey of Data Pricing for Data Marketplaces}.
\newblock \bibinfo{journal}{\emph{IEEE Transactions on Big Data}} (\bibinfo{year}{2023}).
\newblock


\bibitem[Zheng et~al\mbox{.}(2017)]%
        {zheng2017overview}
\bibfield{author}{\bibinfo{person}{Zibin Zheng}, \bibinfo{person}{Shaoan Xie}, \bibinfo{person}{Hongning Dai}, \bibinfo{person}{Xiangping Chen}, {and} \bibinfo{person}{Huaimin Wang}.} \bibinfo{year}{2017}\natexlab{}.
\newblock \showarticletitle{An overview of blockchain technology: Architecture, consensus, and future trends}. In \bibinfo{booktitle}{\emph{2017 IEEE international congress on big data (BigData congress)}}. Ieee, \bibinfo{pages}{557--564}.
\newblock


\bibitem[Zichichi et~al\mbox{.}(2020)]%
        {zichichi2020ensuring}
\bibfield{author}{\bibinfo{person}{Mirko Zichichi}, \bibinfo{person}{Contu Michele}, \bibinfo{person}{Stefano Ferretti}, \bibinfo{person}{Rodr{\'\i}guez-Doncel V{\'\i}ctor}, {et~al\mbox{.}}} \bibinfo{year}{2020}\natexlab{}.
\newblock \showarticletitle{Ensuring personal data anonymity in data marketplaces through sensing-as-a-service and distributed ledger}. In \bibinfo{booktitle}{\emph{CEUR Workshop Proceedings}}, Vol.~\bibinfo{volume}{2580}. Sun SITE Central Europe/RWTH Aachen University, \bibinfo{pages}{1--16}.
\newblock


\bibitem[Zyskind et~al\mbox{.}(2015)]%
        {zyskind2015decentralizing}
\bibfield{author}{\bibinfo{person}{Guy Zyskind}, \bibinfo{person}{Oz Nathan}, {et~al\mbox{.}}} \bibinfo{year}{2015}\natexlab{}.
\newblock \showarticletitle{Decentralizing privacy: Using blockchain to protect personal data}. In \bibinfo{booktitle}{\emph{2015 IEEE security and privacy workshops}}. IEEE, \bibinfo{pages}{180--184}.
\newblock


\end{thebibliography}


\end{document}